\documentclass[12pt,aps,nofootinbib,showpacs,preprint,preprintnumbers]{revtex4}

\usepackage{epsfig,subfigure}

\newcommand{\Dslash}{\not{\hbox{\kern-4pt $D$}}}

\def\beq{\begin{equation}}
\def\eeq{\end{equation}}
\def\bea{\begin{eqnarray}}
\def\eea{\end{eqnarray}}

\begin{document}

\preprint{\vbox { \hbox{WSU--HEP--0805} \hbox{MCTP-08-44}
\hbox{NSF-KITP-0861} }}
\title{\boldmath Initial determination of the spins of the gluino and
  squarks at LHC}
\author{Gordon L. Kane$^1$, Alexey A. Petrov$^{1,2}$, Jing Shao$^1$
  and Lian-Tao Wang$^3$}
\affiliation{$^1$Michigan Center for Theoretical Physics, University
  of Michigan, Ann   Arbor, Michigan 48109 \\
$^2$Department of Physics and Astronomy, Wayne State University, Detroit,
Michigan 48201\\
$^3$Physics Department, Princeton University, Princeton, New
Jersey 08544}
\date{\today}

\begin{abstract}
In principle particle spins can be measured from their production
cross sections once their mass is approximately known. The method
works in practice because spins are quantized and cross sections
depend strongly on spins. It can be used to determine, for
example, the spin of the top quark. Direct application of this
method to supersymmetric theories will have to overcome the
challenge of measuring mass at the LHC, which could require high
statistics. In this article, we propose a method of measuring the
spins of the colored superpatners by combining rate information
for several channels and a set of kinematical variables, without
directly measuring their masses. We argue that such a method could
lead to an early determination of the spin of gluino and squarks.
This method can be applied to the measurement of spin of other new
physics particles and more general scenarios.
\end{abstract}

\vspace*{1in} \pacs{}

\maketitle



\section{Introduction}

The upcoming experiments at the Large Hadron Collider (LHC) will
shed light on the solutions of the shortcomings of the Standard
Model (SM), and should identify new degrees of freedom associated
with the New Physics. Most of the leading candidates for such
physics involve supersymmetry, which predicts doubling of the
observed particle spectrum. It is expected that some of the
supersymmetric partners of the SM particles will be produced at
the LHC. It will be essential to confirm that the candidate
superpartners indeed differ by half a unit in spin from their
partners in order to be confident it is indeed supersymmetry that
is being discovered. A rigorous determination of the spins of new
particles will be difficult and complicated, particularly early
with limited statistics, and will require a detailed analysis of
decay angular distributions and correlations
\cite{Barr:2004ze}-\cite{Alves:2006df}. Though it will be
necessary to do this eventually, we want to argue that an initial
determination of the spins can in many cases be made, particularly
for particles that dominate production. If the new physics is
indeed supersymmetric the initial production of colored
superpartners is likely (though not certain) to dominate, and we
focus on a preliminary measurement of their spins. As an
illustration of the method, we apply it to the determination of
the spin of the top-quark, whose measurement by this approach does
not seem to have been previously reported. The top-quark analysis
illustrates both the strength and limitations of the method,
clearly distinguishing a spin-$\frac{1}{2}$ ``top quark" from a
spin-0 one, but not from a spin-1 one.

A standard way of testing a theory, such as supersymmetry, is by
comparing theoretical predictions with experimental data. Given
preliminary observation of a signal for a new particle, the
minimal test would simply involve calculation of the production
rate for a particle of a given spin, color, and other quantum
numbers, and checking that the theory and data are consistent. For
instance, the production rates of colored superpartners, like that
of the top quarks, are largely determined by their QCD
interactions, so these checks do not depend significantly on other
features of the chosen model of physics beyond the SM.

Such a test can only be performed once the particle's mass is
measured, since the production rate depends on the mass. In order
to test a particular model, one can in addition compare the
production rates computed within this model with that of some
alternative scenario -- comparisons are normally less sensitive to
systematic uncertainties such as cross section normalization
errors. For example, for the top quark, one can compute the rate
for the particle in the same representation of color group, but of
different spin, say zero or one. These rates would differ
significantly both because of the number of spin degrees of
freedom and because of different angular momentum configuration of
the final state; depending on their spin, the final state
particles can be produced in either $s-$ or $p-$wave. As it turns
out, the latter effect is significant even at a hadron collider.
As we shall see, even an approximate measurement of the mass
provides good discrimination between the cases of different spin,
simply because the resulting production rates are very different.
This implies that one can determine the spin of the produced
particles even with the limited-luminosity early measurements of
production cross sections and mass. The results can of course be
improved with more data and analysis. Similarly, we will compare
alternative spin scenarios for  color octet ``gluinos". The method
can later be extended to test both the color and spin separately.

Of course accurate calculations of cross sections of
strongly-interacting particles require inclusion of higher order
QCD effects. In order to demonstrate the effectiveness of the
method, here we only consider the rates computed at the tree
level. Indeed, the choice of the renormalization and factorization
scales, as well as other effects, will bring additional
uncertainty into evaluation of the absolute rates. For a given
mass, the effects of such uncertainties will be essentially the
same for the different spin cases. However, as we shall show in
this paper, the relevant comparison is between different spin
scenarios with {\it different} mass scales. Therefore, the effect
of higher order QCD corrections do not cancel in this comparison.
Nevertheless, we shall argue that such corrections will not change
the rate in ways that significantly affect the effectiveness of
our method.

It can be challenging to measure the mass of a particle,
particularly at a hadron collider. Most importantly, typical
kinematical variables, such as $H_T$ or $m_{\rm eff}$, essentially
measure the {\it mass differences} between the produced particle
and its decay chain members. Absolute mass scale can seldom be
determined from these variables in scenarios such as
supersymmetry, where the end product of a decay chain is a massive
neutral stable particle which escapes detection. With only
information about rates in certain channels, and mass differences,
as we will discuss in detail below, there are typically
degeneracies between supersymmetry and other scenarios with
different spin partners.

Studies of mass determination using detailed kinematical
information were performed in
\cite{Hinchliffe:1996iu,Bachacou:1999zb,Allanach:2000kt,Gjelsten:2004ki,Gjelsten:2005aw,Lester:2006yw,Lester:2005je,Ross:2007rm,Barr:2007hy,Cho:2007dh,Lester:2007fq,Cheng:2007xv,Cheng:2008mg,Tovey:2000wk}.
Application of these methods typically requires rather high
statistics. Moreover, such mass measurement would have to be
fairly accurate in order to be useful for spin determination.
Since the rate scales with a high power of mass, a small error in
mass measurement could translate into a large uncertainty in rate
predictions. In this paper, we suggest a rather different method.
Instead of trying to measure the mass directly, we propose to
combine the rate of several channels and a set of commonly used
kinematical distributions. We demonstrate that this method allows
us to break the degeneracy and could lead to an early
determination of the spins of the new particles produced at the
LHC.

Ideally, one would also like to measure gauge couplings, color
charges, and electroweak charges of the new particles, which are
also predicted by the theory, and also test spins of other
particles (e.g. charginos and neutralinos) that are produced with
smaller rates and/or in decay chains. Such rates and branching
ratios do depend on particle spins, so once the spins and masses
of the particles that are dominantly produced are determined, some
tests or determinations of others may be possible, depending on
how well the relevant signals can be isolated. Some constraints on
color charges can be checked fairly simply. For example, if
gluinos are produced from color octet gluons, their color
representation can be $8\otimes 8=1+ 8+ 8 + 10 + \overline{10} +
27$, and one can calculate the production cross sections holding
other quantum numbers fixed. It is appropriate to proceed by
initially checking one property at a time (such as spin), setting
others (such as color) at their expected values. As more
statistics become available, several variables can be
simultaneously varied.

This paper is organized as follows: we motivate our discussion by considering measurement of the
top-quark spin in Sec. II, provide basic set-up in Sec. III and apply the proposed method to determination
of the spin of gluino in Sec. IV. We conclude in Sec. V.

\section{Motivation: determining the spin of the top-quark}

\begin{figure}[tb]
\centerline{
\includegraphics[width=4.5in,angle=0]{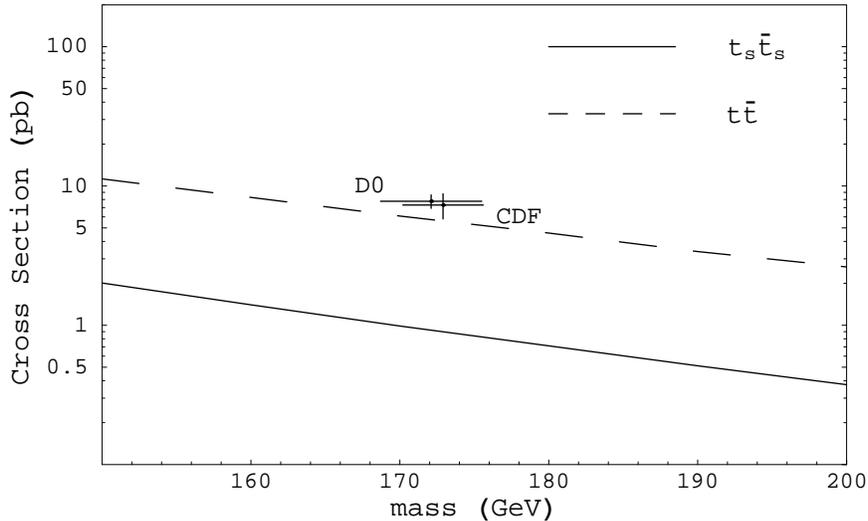}}
\caption{Cross section as a function of the ``top quark" mass at
the Tevatron. The lower dashed line is for the case of
spin-$\frac{1}{2}$ ``top quark", and the solid line is for the
case of spin-0 ``top quarks". The two points with error bars show
the CDF and D0 data for the top mass and cross section with one
$\sigma$ error
\cite{Tevatron-top-rate-cdf,Tevatron-top-rate-d0,Tevatron-top-mass-cdf,Tevatron-top-mass-d0}.
} \label{TopFig}
\end{figure}

In this section, we briefly discuss the method of determining the
spin of a candidate top-quark, denoted ``top-quark," with
production rate information. The production of the top quark at
the Tevatron is dominated by the $s-$channel process $q\bar
q\rightarrow t\bar t$, which implies that the cross section is
proportional to the velocity $\beta$ of the produced top-quark
near the kinematical threshold. However if instead scalar
particles are produced, the configuration of the final state must
be CP even and so the cross section is dominantly $p-$wave and is
proportional to $\beta^3$ near the threshold. Thus we expect the
spin-$\frac{1}{2}$ ``top quark" to have a larger production rate
than spin-0. In Fig.\ref{TopFig}, we plot the cross sections as
functions of mass of the ``top quark", where we also included the
gluon fusion process. The production cross-section for spin-0
``top quarks" is indeed significantly below that for the
spin-$\frac{1}{2}$ ``top quark". Thus given the mass of the ``top
quark", the magnitude of the production cross-section strongly
favors the case of spin-$\frac{1}{2}$ ``top quark"\footnote{Indeed, 
it should be remarked that if a top-quark candidate decays into a
SM b-quark and a W boson, it can only have a half-integral spin.}.

\section{Cross section for colored partners with different spin}

In this section, we compare the production cross sections of
colored superpartners and their counterparts (with different spin)
in alternative scenarios.

For the purpose of illustrating the idea, we implemented a simple
same-spin scenario\footnote{One should note that this is not the
case of the Universal Extra-Dimension (UED)
scenario\cite{Appelquist:2000nn}.} where for each SM quark $q$
there is a massive partner $q'$. Results for models with more
extended fermion partner sectors could be obtained by scaling from
our result. The coupling relevant for production is schematically,
\beq \label{same-spin} {\cal L}_{int}={\bar q}'_L
\displaystyle{\not} g\,' q_L + {\bar u}'_R \displaystyle{\not}g\,'
u_R + {\bar d}'_R \displaystyle{\not}g\,' d_R\eeq

\begin{figure}[tb]
\centerline{
\includegraphics[width=4.5in,angle=0]{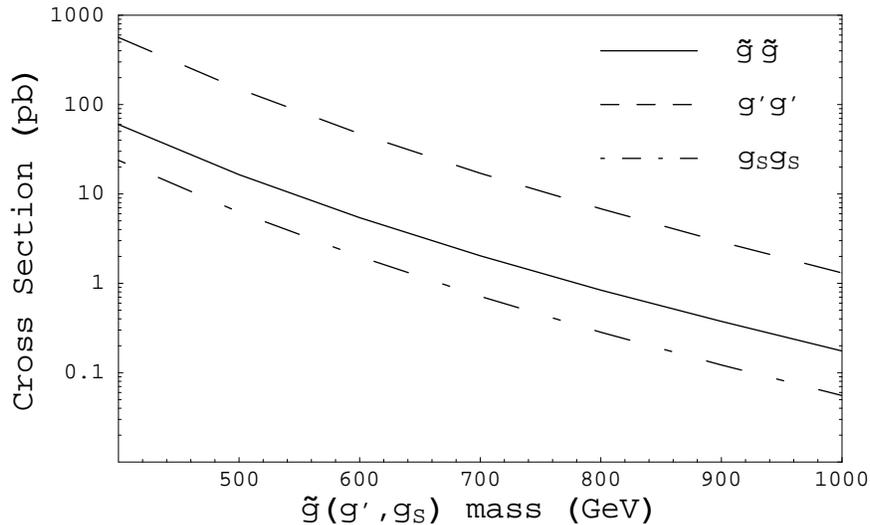}}
\caption{A plot of the cross sections for gluino pair production
(solid line), spin-1 gluon partner $g'$ pair production (dashed
line) and spin-0 gluon partner $g_S$ pair production (dot-dashed
line) at LHC. In the calculation, extra color triplets (e.g.
$\tilde q$ or $q'$) are taken to be $5$ TeV.
} \label{GluinoFig}
\end{figure}

We begin with spin $0$ ($g_S$)\footnote{We consider a real scalar here.}, $\frac{1}{2}$ ( $\tilde g$) and
$1$ color octets ($g'$). The QCD interactions of these color
octets with the quarks and gluon are completely fixed by gauge
invariance. The parton cross sections for pair production of these
particles are already calculated in
reference~\cite{OctetScalars,OctetVectors,Smillie:2005ar,Dawson:1983fw,Beenakker:1996ch}.
For example, at the LHC, gluon-gluon fusion leads to the following
elementary cross-sections
~\cite{OctetScalars,OctetVectors,Smillie:2005ar,Dawson:1983fw,Beenakker:1996ch},
\begin{eqnarray}
\hat \sigma_{g g \to g_S g_S}(\hat s, m_{g_S}) &=& \frac{\pi
\alpha_s^2}{\hat s} \left[ \left(\frac{15}{16} +
\frac{51m_{g_S}^2}{8 \hat s}\right)\beta + \frac{9 m_{g_S}^2}{2
\hat s^2} \left(\hat s - m_{g_S}^2 \right)
\log\frac{1-\beta}{1+\beta} \right],\\
\hat \sigma_{g g \to \tilde g \tilde g}(\hat s, m_{\tilde g}) &=&
\frac{\pi \alpha_s^2}{\hat s} \left[ - \left(3+\frac{51 m_{\tilde
g}^2}{4\hat s} \right) \beta+\frac{9}{4} \left(1+\frac{4 m_{\tilde
g}^2}{\hat s}-\frac{4 m_{\tilde g}^4}{\hat s^2} \right)
\log\frac{1+\beta}{1-\beta} \right],\\
{\hat \sigma}_{gg\rightarrow g_V g_V}(\hat s,
m_{g'})&=&\frac{\pi\alpha_s^2}{\hat s}\left[\left(9 \frac{\hat
s}{m_{g'}^2} + \frac{117}{8} +
\frac{153}{4}\frac{m_{g_V}^2}{{\hat s}}\right)\beta +9\left(1 + 3
\frac{m_{g'}^2}{\hat s} - 3\frac{m_{g_V}^4}{{\hat s}^2}\right)
\log\frac{1 - \beta}{1 + \beta}\right]
\end{eqnarray}
These results are obtained including diagrams with $s-$channel
gluon fusion and $t-$channel gluino exchange. For the spin-0 and
spin-1 cases, there is also a four-point interaction diagram. From
the above result, we see that they have the same functional
dependence on $\beta$, so the argument applied for the top-quark
case does not apply here. However the differences in the cross
sections originate from the differences in the interaction form
determined by the spin structure. As for the top quark case, one
again finds the longitudinal enhancement for the cross section of
the spin-1 gluino. In Fig.\ref{GluinoFig}, cross sections at LHC
of the three cases are plotted as functions of the mass. In our
calculation, the factorization scale $\mu_F$ and renormalization
scale $\mu_R$ are set to be $\mu_F = \mu_R = m_{\tilde{g}, g',
g_S}$. From the plot, we can see that the cross section increases
by roughly an order of magnitude as spin increases from $0$ to $1$
for a given mass. On the other hand, these cross sections decrease
rapidly with mass. Roughly speaking, for a increase in mass by
$200$-$300$ GeV, the cross section decreases by an order of
magnitude. Thus to match a given result of the cross section, the
masses of gluino for different spin must be different by
$200$-$300$ GeV. This makes it possible to determine the spin by a
rough mass measurement.

\begin{figure}[tbp]
\begin{center}
\includegraphics[width=3.6in,angle=0]{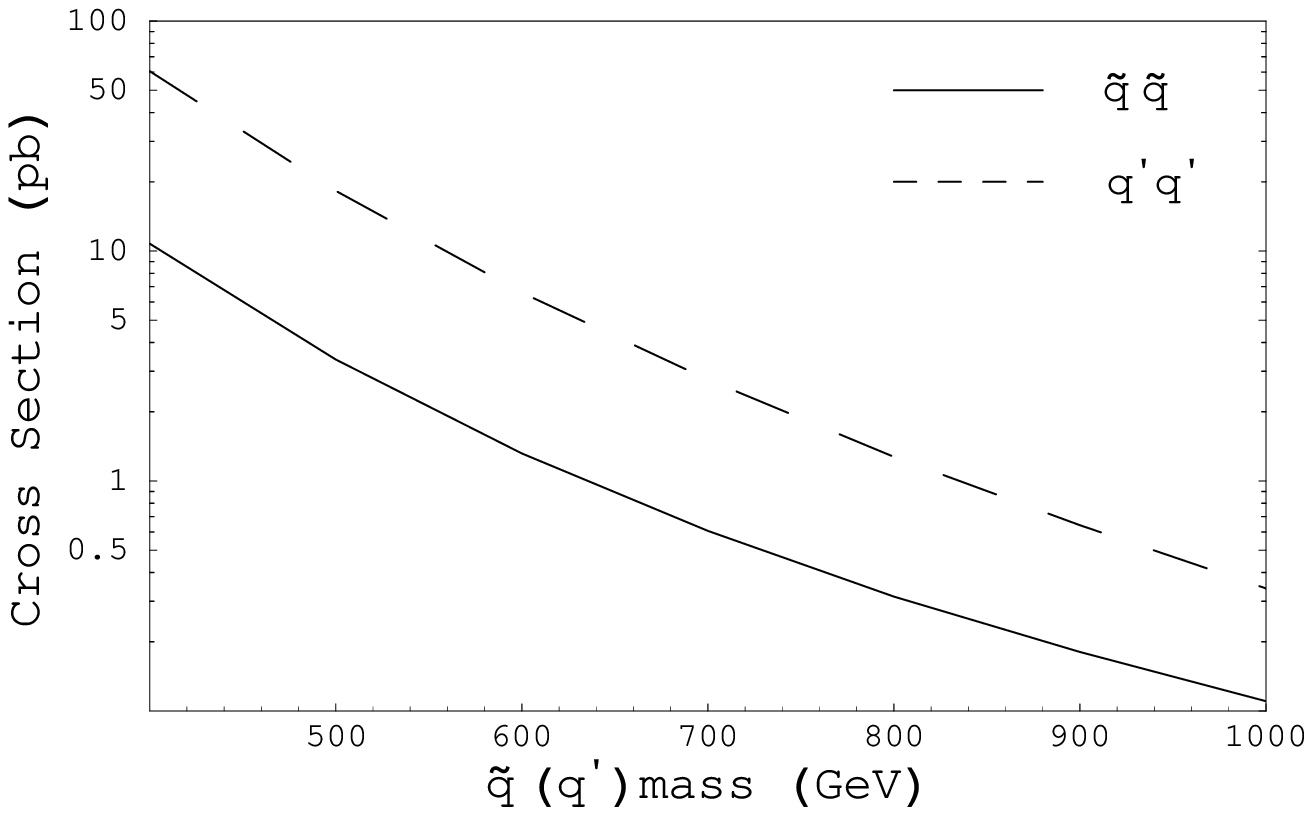}
\includegraphics[width=3.6in,angle=0]{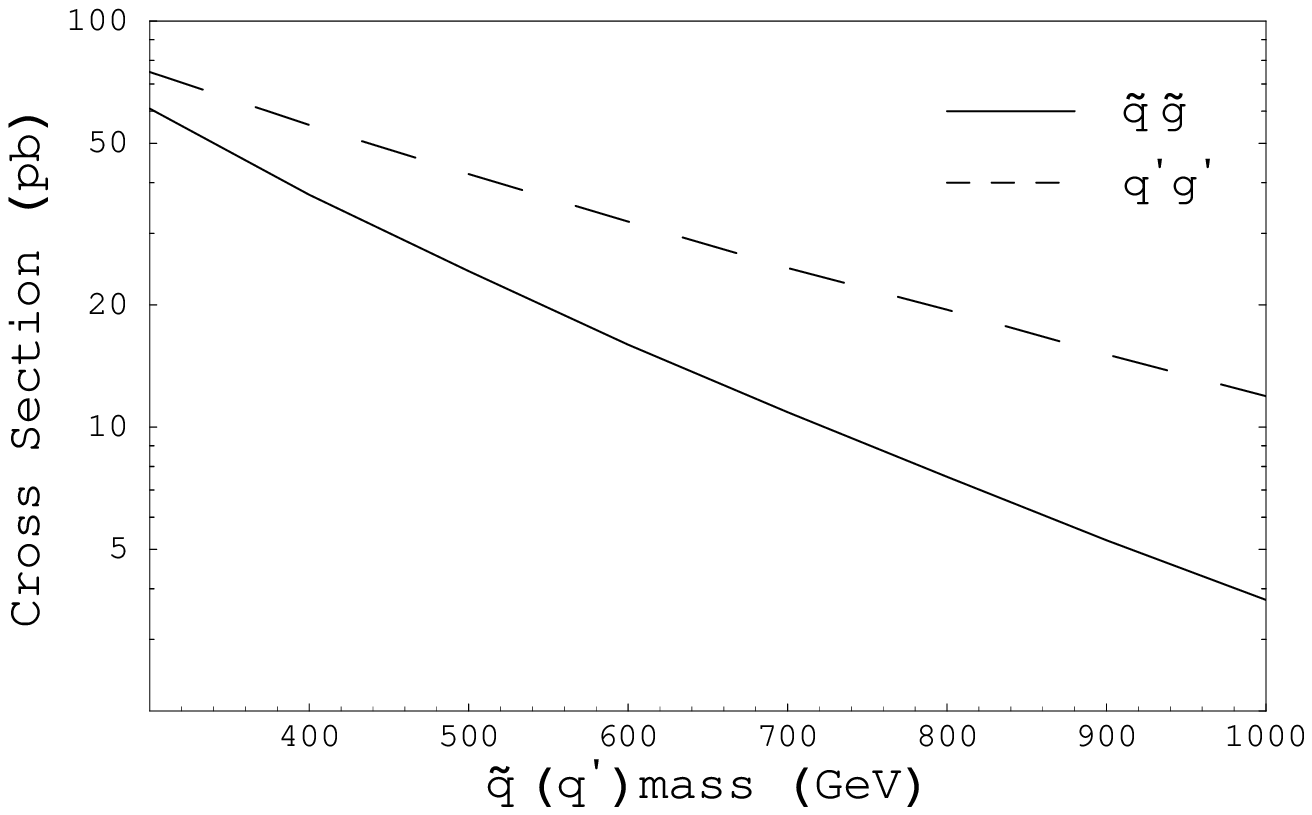}
\caption{Top: Cross sections for squark pair production (solid
line) and fermionic quark partner $q'$ pair production (dashed
line) at LHC. In the calculation, gluino and $g'$ are taken to be
$5$ TeV. Bottom: Cross sections for
squark-gluino associated production (solid line) and $q'$- $g'$
associated production (dashed line) at LHC. The gluino mass and
$g'$  mass are $608$ GeV and $850$ GeV respectively in order to
match the cross sections of gluino pair production and $g'$ pair
production. } \label{Fig:squark-gluino}
\end{center}
\end{figure}

For later reference, we also present here the numerical results
for squark pair (fermionic partner $q'$ pair) production and
squark-gluino ($q'$-$g'$ ) associated production in
Fig.~\ref{Fig:squark-gluino}. For the associated production, the
cross section depends on both the squark ($q'$) mass and the
gluino ($g'$) mass. To give a simple example and for later
convenience, we choose gluino mass and $g'$ mass such that their
pair production rate are matched. Then we plot the cross section
as a functions of squark ($q'$) mass.

The cross sections are calculated using MadGraph/MadEvent
\cite{madgraph}. In the calculation, we use the
parton distribution function CTEQ6L \cite{cteq6}. The renormalization scale
$\mu_R$ and the factorization scale $\mu_F$ are chosen
$\mu_R=\mu_F= m_{\tilde{q}, q'}$ and $\mu_R=\mu_F= 0.5\times (
m_{\tilde{q}, q'} + m_{\tilde{g},g'})$
for squark (or $q'$) pair production and squark ($q'$) gluino ($g'$)
associated production, respectively.

Our calculation of the parton cross sections is done at
tree-level. The next-to-leading order calculation usually changes
the result by a factor of order unity
\cite{nlo_gluino_squark,Plehn:2005cq,nlo_stop}. As the
multiplicative K-factors always increase the production cross
section, such corrections should not change the overall hierarchy
of the rates for different spin assignments. The enhancement
factors of the cross section for particles with different spin but
in the same color representation have the same color factor and
the same initial state. Although not identical, those factors are
not expected to be significantly different. Notice also we are not
comparing cross sections at the same mass scale. As we will
discuss in detail below, we will match the cross sections of
particles with different spin with different masses. However,
those different masses are typically within a factor of 2.
Therefore, additional effects from different choices of
renormalization and factorization scale will have to be included.
We note that such differences in K-factors are typically quite
small, about $10 \%$ \cite{nlo_gluino_squark,nlo_stop} across the
mass range we are interested in. We have also partially taken this
into account by chosing $\mu_R $ and $\mu_F $ to be correlated
with the masses of the particle produced.

\section{Spin Determination}

In this section, we present our study of spin determination for
the possible cases where degeneracy could occur. First, we briefly
review and clarify the degeneracy by combining rate and mass
differences. For concreteness, we will focus on the comparison
between supersymmetry and the  same spin scenario defined in
Eq.~\ref{same-spin}. We will assume the final decay products
always include stable neutral particles. Of course, this would be
the LSP in the case of SUSY. We will assume the existence of such
a stable particle, denoted by $A$, in the case of same spin
scenario due to the implementation of certain discrete symmetry.
Examples which have qualitatively the same feature as such a
scenario include Universal Extra-Dimensions
\cite{Appelquist:2000nn} and little Higgs models with T-parity
\cite{Cheng:2003ju,Cheng:2004yc}.

For simplicity, we will only consider simple decay chains
following the production. A study of a set of models with more
complicated decay chains is beyond the scope of this initial
paper. We will briefly comment on such scenarios in the
conclusion.

\subsection{Degeneracy with rate and mass differences}

As pointed out in Ref.~\cite{Cheng:2005as} and
\cite{Meade:2006dw}, typical transverse variables are largely
sensitive to the mass differences between the initial and
subsequent particles in the decay chain. This is also a physics
reason for the set of degeneracies in measuring mass parameters
within supersymmetry, pointed out in
Ref.~\cite{ArkaniHamed:2005px}. Therefore, in a particular
channel, there can be a degeneracy between supersymmetry and an
alternative scenario with the same-spin partners. Schematically,
suppose the production cross section for some superpartner of mass
$m$ is $\sigma^{\rm SUSY}_{\rm m}$. We could choose the mass of a
similar partner in the alternative scenario, $m'$, so that the
production cross section matches, $\sigma_{m'}^{\rm altn} =
\sigma^{\rm SUSY}_{m}$. At the same time, we could learn from some
transverse kinematical variables the information about mass
differences in supersymmetry, say $\Delta m = m-m_{\rm LSP}$.
However, if those variables mainly contain information about mass
differences (which is true to a pretty good approximation), we can
adjust the mass difference in the alternative scenario, $\Delta m'
= m'-m_{A}$, to fix the kinematical distribution as well. If other
information is available about the mass, that can remove the
degeneracy, but typically that will not occur with early collider
data.

We demonstrate this degeneracy in a simple example. Consider the
comparison between a squark and some fermionic quark partner $q'$.
Let's also assume they are both pair produced, and followed by
similar simple decay chains $\tilde{q} \rightarrow q + {\rm LSP}$
and $q' \rightarrow q + A$, where $A$ is some stable neutral
particle, analogous to the lightest supersymmetric particle (LSP)
in SUSY models. In our simulation, we generate parton-level events
with MadGraph/MadEvent \cite{madgraph} for the squark pair
production and decay $pp\rightarrow \tilde{q} \tilde
{q}^*\rightarrow q\bar q + 2\,\rm LSP$, as well as the $q'$ pair
production and decay $pp\rightarrow q'q'^*\rightarrow q\bar q A
A$ \footnote{Here, for simplicity, we haven't include the $\tilde q\tilde q$
and $q'q'$ final states, and only include the quark partners of $u_R$ in the final state.}.
Then the parton-level events are passed to PYTHIA 6.4
\cite{pythia6} for parton shower, fragmentation, and PGS4
\cite{pgs4} for detector reconstruction. It is straightforward to
match the production rate by choosing appropriate $m_{\tilde{q}}$
and $m_{q'}$. For example, $m_{\tilde q}=549$ GeV and $m_{q'}=900$
GeV both give rise to $\sigma\approx 2$~pb. In the above examples,
we have not decoupled the color octet particles, but this will not
affect our discussion below.

\begin{figure}[tbp]
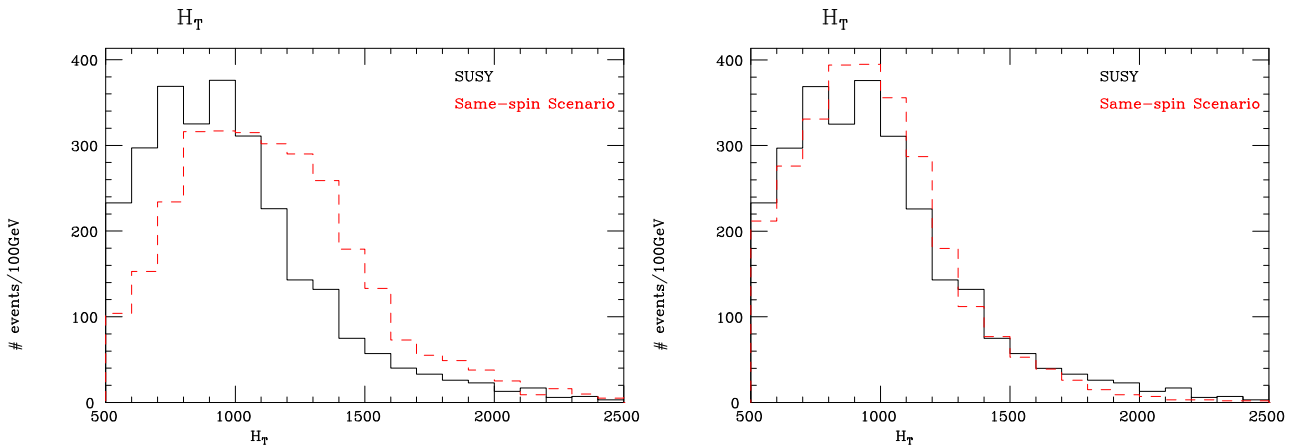

\begin{center}
\includegraphics[width=2.3in,angle=90]{figs/uruk_ht.eps}
\includegraphics[width=2.3in,angle=90]{figs/uruk_ht1.eps}
\caption{Left: $H_T$ distributions for squark pair production
(solid line) and $q'$ pair production (dashed line) at LHC with
both the cross section and mass difference matched. The mass
parameters involved are $m_{\tilde q}=549$ GeV, $m_{q'}=900$ GeV,
$m_{\rm LSP}=97$ GeV and $m_{A}=448$ GeV. Right: $H_T$
distributions for the same processes but with the mass difference
adjusted in the same spin scenario. The mass of the missing
particle now is $m_A=548$ GeV. In both cases, we have generated 3000 events, 
corresponding to approximately $1.5$ ${\rm fb}^{-1}$ integrated luminosity.} \label{Fig:uruk-ht}
\end{center}
\end{figure}

Next, we adjust the mass of the missing particle to match the
kinematical distributions. Notice that choosing the mass
difference to be the same, $\Delta m = m_{\tilde{q}}-m_{\rm LSP} =
m_{q'} - m_{A}$, will give rise to significant difference in
kinematical distribution, particularly when the mass difference is
large. Indeed, in order to match the rate, we have to choose
$m_{q'}$ to be significantly greater than $m_{\tilde{q}}$.
Therefore, choosing a large and fixed mass difference usually
implies  $\Delta m \gg m_{\rm LSP}$. In this case, the available
energy $\sim \Delta m $ tends to be shared evenly between the LSP
and the rest of the observable decay products ($q$ in this case).
On the other hand, in the alternative scenario, we usually have
$\Delta m $ much closer to the missing particle mass $m_{A}$. It
is therefore typical for the recoiling quark to have a harder
spectrum. In the Fig.\ref{Fig:uruk-ht}, we plot the $H_T$
\footnote{Effective mass is defined as $H_T=\displaystyle{\not}E_T
+ \sum_{all jets} P_T^a$} distribution for $\tilde q$ and $q'$
pair productions. In both cases, we keep the same mass difference
$\Delta m=452$ GeV, i.e., $m_{\rm LSP}=97$ GeV and $m_{A}=448$
GeV. One can check that the difference in the average of the $H_T$
does match correctly with the theoretical expectation.

However, this does not imply the absence of the degeneracy. The
key requirement for the existence of such a degeneracy is that the
transverse variables are approximately functions of {\it only}
mass differences, even if the functions may be different for
different scenarios. In this case, it is still possible to adjust
the mass difference in both scenarios to obtain indistinguishable
distributions, even though the mass differences will be different
in those cases. Although this requires very specific masses, it is
important that we address this possibility in order to have a
robust spin determination method. We can see this again in the
example of squark production and decay. For the examples we just
considered, if $M_A$ is increased to $548$ GeV, then one cannot
distinguish the $H_T$ of these two cases as seen in the
Fig.\ref{Fig:uruk-ht}.

\begin{figure}[tb]
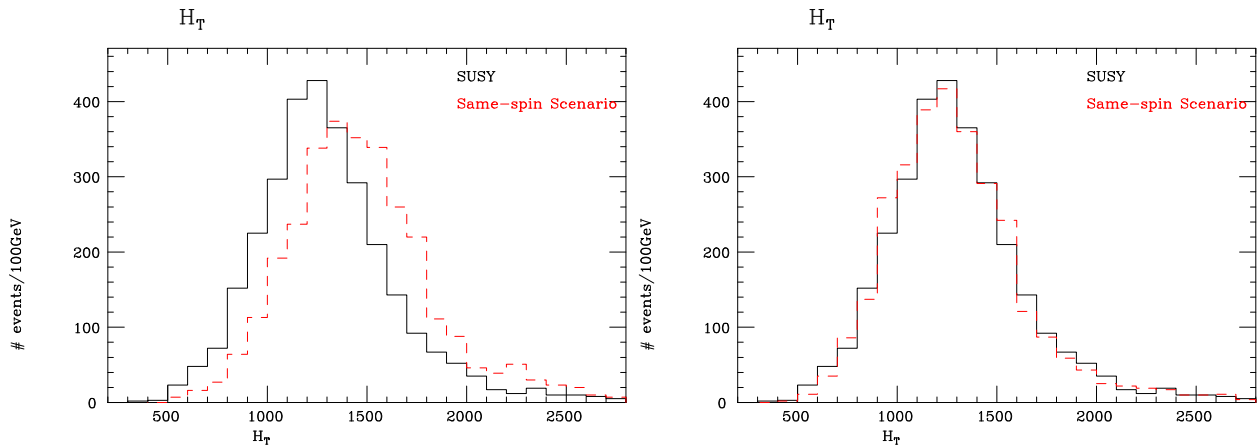

\begin{center}
\includegraphics[width=2.3in,angle=90]{figs/gogv_ht.eps}
\includegraphics[width=2.3in,angle=90]{figs/gogv_ht1.eps}
\caption{Left: $H_T$ distributions for gluino pair production
(solid line) and $g'$ pair production (dashed line) at LHC with
both the cross section and the mass difference matched. The mass
parameters involved are $m_{\tilde g}=800$ GeV, $m_{q'}=1060$ GeV,
$m_{\rm LSP}=137$ GeV and $m_{A}=397$ GeV. Right: $H_T$
distributions for the same processes but with the mass difference
adjusted in the same spin scenario. The mass of the missing
particle is now $m_A=497$ GeV. In both cases, we have generated 3000 events,
corresponding to approximately $3.75$ ${\rm fb}^{-1}$ integrated luminosity.} \label{Fig:gogv-ht}
\end{center}
\end{figure}

A similar conclusion applies to three body decays such as
$\tilde{g} \rightarrow q \bar{q} + {\rm LSP}$. More explicitly, we
consider supersymmetric and same-spin partners, which have about
the same production rate ($\sigma\approx 0.8$~pb) and mass
splitting,
\begin{itemize}
\item $m_{\tilde g}=800$ GeV, $m_{\rm LSP}=137$ GeV \item
$m_{g'}=1060$ GeV, $m_{A}=397$ GeV.
\end{itemize}
The $H_T$ distributions for both productions are plotted in Fig.
\ref{Fig:gogv-ht}. Again we can adjust the mass difference in the
same spin scenario such that the two distributions match. This can
be seen in Fig.\ref{Fig:gogv-ht}, where we have increased $m_{ A}$
to $497$ GeV.

Before proceeding further, we remark that there is a possibility
that we could tell that the partners are from the same spin
scenario and not supersymmetry. Assume the new physics discovered
at the LHC is actually in the same spin scenario. We want to
verify whether a supersymmetric scenario could match the rate and
the same kinematical distributions. Because of the difference in
production rate, the supersymmetric scenario will necessarily have
a much lighter colored partner, squark or gluino, or both. Then,
in order to macth the $H_T$ distribution, we must have at least
comparable mass gaps between the colored partner and the missing
particle. Therefore, there is a possiblity in this case that we
cannot find a solution for the supersymmetric scenario. For
example, from the previous discussion, we know that, to match a
cross section about $0.8$~pb, either a gluino with mass $m_{\tilde
g}=800$ GeV or a $g'$ with mass $m_{g'}=1060$ GeV is required.
However, if the mass difference is measured to be larger than
$700$ GeV, then the supersymmetric scenario is excluded since
otherwise the LSP mass is below the LEP II bound.

\subsection{Matching multi-channels}

As we have seen so far, there are purely kinematical differences
for models with different mass of the decaying particle. We also
notice that these differences are generally not very significant
for measuring spins since the mass difference $\Delta m$ can be
adjusted independently of the mass of the decaying particle. A
more detailed analysis for these situation is certainly necessary
to explore such differences for  spin determinantion. However, in
reality, there will be more handles to distinguish spin using the
rate information. A complete model usually contains several other
new particles beside the color octet In many theories, we expect
the  masses of the colored partners to be similar, i.e., around
TeV scale. Therefore, at least several of them will be copiously
produced at the LHC.  We should consider the production channels
of those colored states together. We will then have more
observables which we could use to break the degeneracy discussed
above. In this section, we present a method to achieve this goal
by combining information from several channels. We show that this
method, based on a set of simple observables, is effective  and
independent of mass measurement.

\begin{figure}[t!]
\begin{center}
\includegraphics[width=4.5in,angle=0]{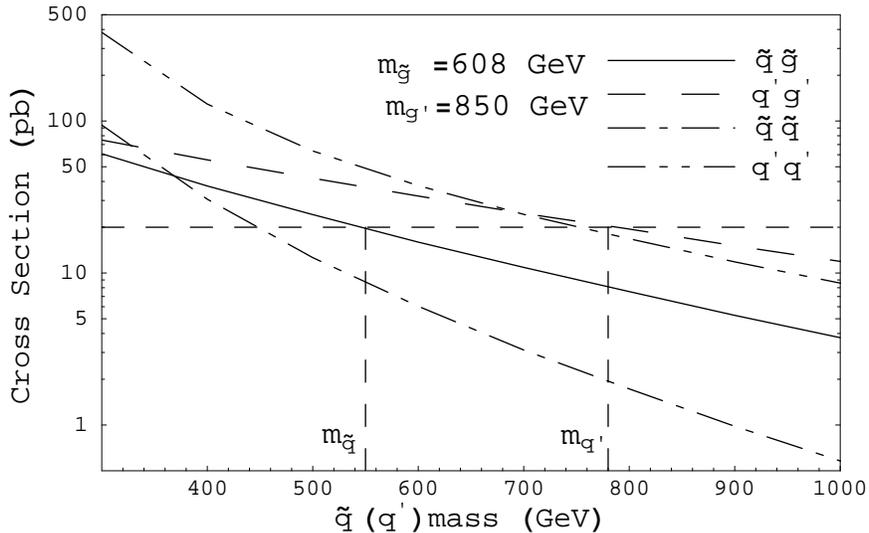}
\caption{Production cross sections for $\tilde g\tilde q$
production (solid line), $g'q'$ pair production (dashed line),
$\tilde q\tilde q$ production (dash-dot line) and $q'q'$
production (dash-dot-dot line). The $\tilde g\tilde g$ and $g'g'$
production cross sections are already matched by choosing gluino
mass $m_{\tilde g}=608$ GeV and $g'$ mass $m_{g'}=850$ GeV. The
horizontal dashed line corresponds to $20$~pb for the cross
section of the associated production. If we match to this rate,
the squark mass is $549$ GeV and the $q'$ mass is $780$ GeV as
indicated by the vertical dashed lines. After this matching, the
cross section of $\tilde q\tilde q$ and $q'q'$ productions are
fixed to be about $8$~pb and $19$~pb respectively.  
}\label{Fig:crossx-multi}
\end{center}
\end{figure}

To demonstrate our method, we focus on distinguishing
supersymmetry and the same spin scenario. We start with the case
in which the squark mass is lighter than the gluino mass; so
gluinos decay through a two-body process. In our example, the
gluino and squark masses are taken to be $m_{\tilde g}=608$ GeV
and $m_{\tilde q}=549$ GeV. Then we find that to match the gluino
pair production rate, the $g'$ mass is fixed to be $m_{g'}=850$
GeV. Furthermore we match the gluino-squark production rate by
varying the $q'$ mass, which leads to $m_{q'}=780$ GeV. This can
be seen from Fig.~\ref{Fig:crossx-multi}, where the two vertical
dashed lines correspond to the masses of $\tilde g$ and $q'$ while
the horizontal dashed line corresponds to $20$~pb for the cross
section of the associated production. After this matching, the
cross sections of $\tilde q\tilde q$ and $q'q'$ productions are
fixed to be about $8$~pb and $19$~pb respectively. The cross
sections for both models are calculated in MadGraph/MadEvent and
are
\begin{eqnarray}
  \textrm{ SUSY}&:& \quad \sigma_{\tilde g\tilde g}\approx 4~{\rm pb}, \quad \sigma_{\tilde g\tilde q}\approx 20~{\rm pb},
  \quad \sigma_{\tilde q\tilde q}\approx 8~{\rm pb}\nonumber\\
  \textrm{Same Spin Scenario}&:& \quad \sigma_{g'g'}\approx 4~{\rm pb}, \quad
  \sigma_{g' q' }\approx 20~{\rm pb},
  \quad \sigma_{q'q'}\approx 19~{\rm pb}
\end{eqnarray}
One can see that the rate of $q'$ pair production cannot be
matched with that of the squark pair production at the same time
as other rates are matched.

More generally, we notice that the cross section of gluino ($g'$)
pair production is almost independent of the squark ($q'$) mass.
Therefore, by matching the cross sections for gluino and $g'$ pair
production, the $g'$ mass is fixed for a given gluino mass. After
that, the gluino-squark ($g'$-$q'$) cross section only changes as
squark ($q'$) mass varies. Fig.~\ref{Fig:crossx-multi} shows a
plot displaying the cross sections as functions of squark ($q'$)
mass with gluino ($g'$) mass fixed.
As we can see from this figure, generally, the three
production rates cannot be matched at the same time. In addition, such
matching will typically force us to consider quite different mass
splittings. Therefore,  we also expect the typical jet transverse
momentum variable, such as $H_T$, will be quite different.

The discussions above on matching individual cross sections are
obviously a simplification to demonstrate the principles. In
reality, after showering and forming the jets, the channels
considered above can give sizable overlapping contributions to the
same final state. Therefore, instead of matching the cross section
for each channel, we perform next a more realistic study by
matching directly to observables, in particular, the jet multiplicity
and $p_T$ distribution (again using $H_T$ as a representative
example). Generally we expect no degeneracy even by considering
the jet multiplicities. The reason is as follows. Since we are including
another particle in the analysis, there are three mass variables
in the models. These three variables can be fixed by matching the
total cross section and the two $H_T$ peaks (if the masses of the
two color particles are considerably different). Because of the
difference in the spins in the two models, the ratios of the cross
sections between different channels are generically different.
Therefore the jet counts are not likely to be the same, because
these channels have different jet-multiplicities.

\subsubsection{Case I: $m_{\tilde g} > m_{\tilde q}$}

Let us start with the case where the gluino is heavier than the
squark in the SUSY model. For simplicity, we take it to be the
SUSY model with mass parameters $m_{\tilde g}=608$ GeV, $m_{\tilde
q}=549$ GeV and $m_{\rm LSP}=97$ GeV. The total cross section of 
the $\tilde g\tilde g$, $\tilde q\tilde q$ and $\tilde g\tilde q$ final states
is approximately $31.6$~pb. Then by varying the mass
parameters $m_{g'}$ and $m_{q'}$ in the same spin scenario, one
can match the total rate of the SUSY model; several choices of
masses are listed in Table \ref{Table:chisq}. For each of these
models, we generate $2000$ events in MadEvent-BRIDGE-Pythia-PGS setup\footnote{
We use BRIDGE\cite{Meade:2007js} to decay the parton events containing 
the gluino($g'$) and squark($q'$) before Pythia and PGS.} and
compare the jet multiplicities with those of the SUSY model. The
decay of the gluino is a two-body process $\tilde g\rightarrow
\tilde q + q$, followed by the decay of squark $\tilde
q\rightarrow q + \rm LSP$. Here we include the first two
generations of squarks for simplicity\footnote{The third
generation squarks($q'$) will decay into b-jets which can be tagged and
studied seperately. }. For the same spin scenario, we consider the
similar decay processes of $g'$ and $q'$: $g'\rightarrow q'+q$ and
$q'\rightarrow q + A$. Once the jet counts are obtained, the
difference between SUSY and same spin scenario can be
characterized by defining a $\chi$-square-like quantity
\begin{eqnarray}
  (\Delta S)^2=\frac{1}{N}\sum_{i=1}^{N}
\left(\frac{n_i^{\rm SUSY}-n_i^{\rm
SameSpin}}{\sigma_i}\right)^2.\label{Eq:chisq}
\end{eqnarray}
In the following, we consider jet counts for $N=5$ bins with
$n_{jet}=1,2,3,4,5+$. Here $n_i^{\rm SUSY}$ ($n_i^{\rm SameSpin}$)
is the number of events in each bin for the SUSY (same spin)
model. The standard deviation is defined by
$\sigma_i^2\equiv(\sigma_i^{\rm SameSpin})^2+(\sigma_i^{\rm
SUSY})^2$, where $\sigma_i=\sqrt{n_i+1}$ for each model. In our
simulation, we have used the PGS4 default detector configuration
and cone jet algorithm with a cone size $0.5$. In addition, we use
the PGS trigger with low thresholds
\begin{itemize}
\item Inclusive $\displaystyle{\not} E_T$ $90$ GeV \item Inclusive
single-jet $400$ GeV \item Jet plus $\displaystyle{\not} E_T$
($180$ GeV, $80$ GeV) \item Accoplanar jet and
$\displaystyle{\not} E_T$ ($100$ GeV, $80$ GeV,  $1< \Delta\phi <
2$) \item Accoplanar dijets ($200$ GeV, $\Delta\phi < 2$)
\end{itemize}
To be realistic, we also impose the selection cuts for jets:
$P_T\ge50$ GeV and $\eta < 2.5$.

\begin{figure}[tb]
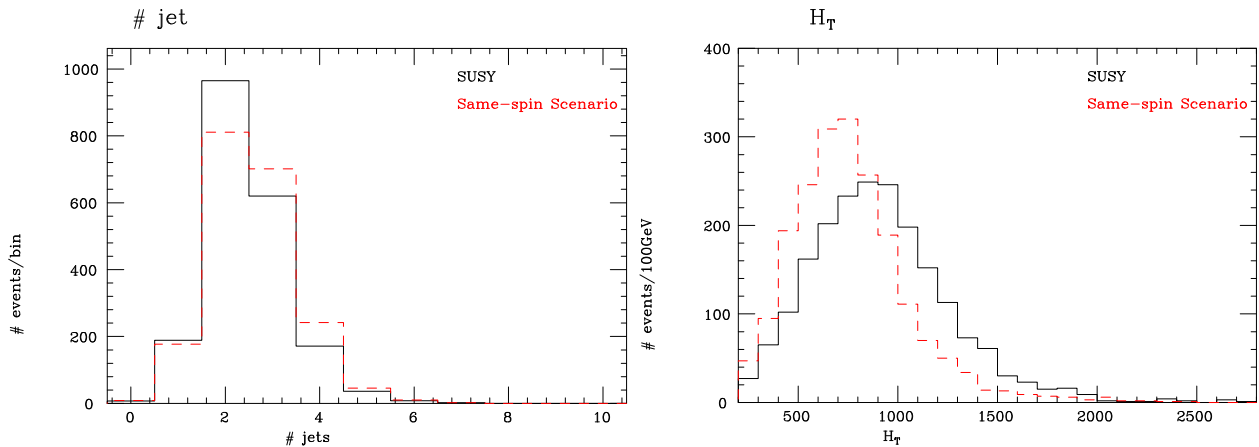

\begin{center}
\includegraphics[width=2.3in,angle=90]{figs/fit1_jet.eps}
\includegraphics[width=2.3in,angle=90]{figs/fit1_ht.eps}
\caption{Left: The jet counts for the SUSY model (solid line) in
the Case I and the same spin model (dashed line) which has the
smallest $\Delta S$ in Table I. We vary mass parameters $m_{\tilde
g}$, $m_{\tilde q}$ and $m_{\rm LSP}$ for the SUSY model, and
$m_{g'}$, $m_{q'}$ and $m_{A}$ for the same spin model. Both
models are tuned to have the same total cross section. Right: The
$H_T$ distributions for the same pair of models: the SUSY model
(solid line) and the same spin model (dashed line). Both plots contain 
2000 events, corresponding to approximately $70$ ${\rm pb}^{-1}$ integrated luminosity.}
\label{Fig:fit1}
\end{center}
\end{figure}

In Table \ref{Table:chisq}, the average of $H_T$ and the value of
$\Delta S$ are shown for given $m_{g'}$, $m_{q'}$ and $m_A$. For
the SUSY model in the comparison, the average of $H_T$ is $913$
GeV, and the jet counts are \{189, 965, 620, 171, 46\} for jet
number $n_j=1,2,3,4,5+$. As can be seen from the table, values of
$\Delta S$ are generically large for cases with $m_{q'}> m_{g'}$.
This can be understood since in this case $q'$ and $g'$ decay very
differently from squark and gluino. However, for cases with
$m_{q'}< m_{g'}$, $\Delta S$ can be smaller. The same spin model
with the smallest $\Delta S$ in the Table is the one with
$m_{g'}=900$ GeV, $m_{q'}=800$ GeV and $m_{A}=550$ GeV, where
$\Delta S=2.5$. The jet multiplicity and $H_T$ distribution are
shown in Fig.\ref{Fig:fit1}.
First of all, one can see that $\Delta S$ is not small, and so the
jet multiplicity is not completely matched.\footnote{It actually
cannot be further reduced by tuning the mass parameters.} Second,
even if $\Delta S$ is small, the $H_T$ distribution is different
enough to eliminates the possible degeneracy. So based on the
above observations, one can see that the degeneracy is unlikely to
exist considering total cross section, $H_T$ and jet multiplicity
simultaneously.

\begin{table}[!ht]
\begin{center}
  \begin{tabular}{|c||c|c|c|c|c|c|}
    \hline
    $m_{A}$ (GeV) & \multicolumn{6}{c|}{Same Spin Models ($m_{g'}$, $m_{q' }$) (GeV) }\\
    \cline{2-7}
    & (1000,~640) &
    (950,~720) & (900,~800)& (850,~880)
    & (800,~960) & (750,~1120) \\ \hline\hline
     100  & (1218,~6.6) & (1271,~7.9) & (1307,~7.2) & (1373,~7.7)  &
     (1399,~13.4)    & (1474,~16.4)  \\ \hline
      250  & (1104,~5.8) & (1178,~6.4) & (1218,~6.2) & (1254,~8.0)  &
      (1289,~16.8)   & (1339,~20.0)  \\ \hline
     400  & (875,~3.2)  & (984,~5.1)  & (1036,~4.8) & (1057,~7.2)  &
     (1072,~17.2)   & (1103,~18.4)  \\ \hline
     550  & (584,~6.5)  & (682,~2.7)  & (767,~2.5)  & (776,~3.9)   &
     (758,~14.6)     & (820,~12.0)    \\ \hline
     700  &  --         & (452,~18.7) & (427,~6.0)  & (410,~7.4)   &
     (456,~6.4)     & (706,~14.2)   \\ \hline
  \end{tabular}
  \end{center}
  \caption{This table shows 
  ($\overline H_T$, $\Delta S$) for each same spin model parameterized by three mass parameters
  $m_{g'}$, $m_{q'}$ and $m_A$ for Case I. Here $\Delta S$ is defined in Eq.(\ref{Eq:chisq}).
  In each column, a set of ($m_{g'}$, $m_{q'}$) is chosen to match the total cross section with that of the SUSY
  model, which is defined by $m_{\tilde g}=608$ GeV, $m_{\tilde q}=549$ GeV and $m_{\rm LSP}=97$
  GeV. In each row, a mass $m_A$ is specified, which can be seen from the first column. For the SUSY model in the comparison,
  $\overline H_T=913$ GeV, and the jet counts are given by \{$189$, $965$, $620$, $171$, $46$\} for jet number
  $n_j=1,2,3,4,5+$. The data indicate that $H_T$ and jet multiplicity of the same spin model cannot
  be matched to those in the SUSY model simultaneously}\label{Table:chisq}
\end{table}
\subsubsection{Case II: $m_{\tilde g} < m_{\tilde q}$}
\begin{figure}[tb]
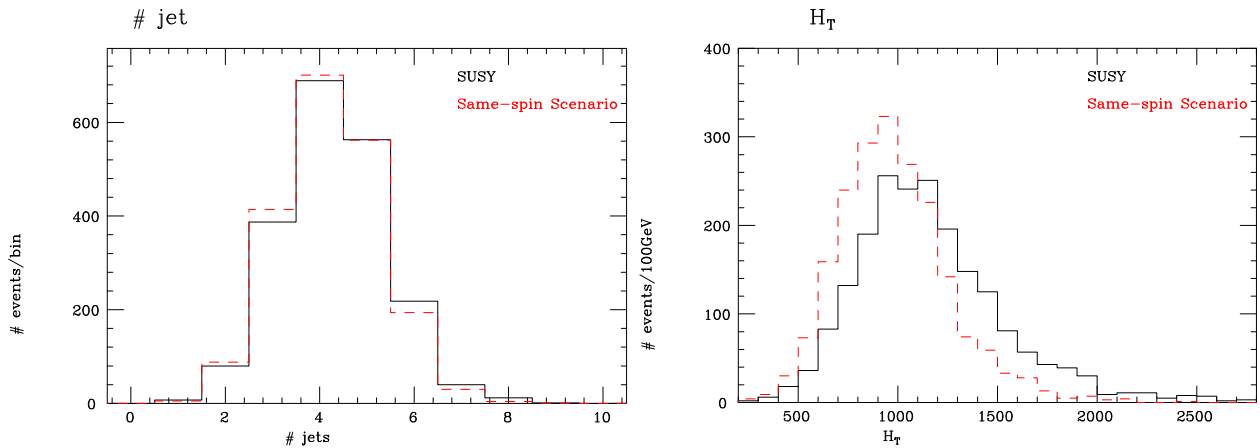

\begin{center}
\includegraphics[width=2.3in,angle=90]{figs/fit2_jet.eps}
\includegraphics[width=2.3in,angle=90]{figs/fit2_ht.eps}
\caption{Left: The jet counts for the SUSY model (solid line) in
the Case II and the same spin model (dashed line) which has the
smallest $\Delta S$ in Table II. We vary mass parameters
$m_{\tilde g}$, $m_{\tilde q}$ and $m_{\rm LSP}$ for the SUSY
model, and $m_{g'}$, $m_{q'}$ and $m_{A}$ for the same spin model.
Both models are tuned to have the same total cross section. Right:
The $H_T$ distributions for the same pair of models: the SUSY
model (solid line) and the same spin model (dashed line). Both plots contain
2000 events, corresponding to approximately $200$ ${\rm pb}^{-1}$ integrated luminosity.}
\label{Fig:fit2}
\end{center}
\end{figure}

For the case where the squark is heavier than the gluino in the
SUSY model, we consider the following example: $m_{\tilde g}=640$
GeV, $m_{\tilde q}=800$ GeV and $m_{\rm LSP}=100$ GeV. The total 
cross section of the $\tilde g\tilde g$, $\tilde q\tilde q$ and 
$\tilde g\tilde q$ final states is approximately $11.3$~pb. In such a
case, the jet multiplicity can be matched better for same spin
models with $q'$ heavier than $g'$. First we vary the masses of
$g'$ and $q'$ to match the total cross section, where four
possible choices of ($m_{g'},m_{q'}$) are listed in Table
\ref{Table:chisq2}\footnote{The reason to choose these four cases
is that $m_{q'}-m_{g'}$ is close to $m_{\tilde q}-m_{\tilde g}$.
For large mass difference, it should have larger kinematical
difference.}. 
The decay processes are simplified by taking $\tilde q\rightarrow
\tilde g + q$ and $\tilde q\rightarrow q+\bar q + \rm LSP$, and
similarly for $q'$ and $g'$. As seen from the table, the best-fit
of jet multiplicity is achieved when $m_{g'}=950$ GeV,
$m_{q'}=1120$ GeV and $m_{A}=600$ GeV, where $\Delta S=0.8$. The
jet multiplicity and $H_T$ distribution are shown in
Fig.\ref{Fig:fit2}. Though the jet multiplicities are well
matched, the differences (both the width and the position of the
peak) in the $H_T$ distributions should be enough to distinguish
them. Therefore, again we see that matching the $H_T$
distributions and matching the jet multiplicities are in conflict
with each other and cannot be satisfied simultaneously. As long as
squarks are not too heavy and the production rate is of about the
same order of magnitude as that of gluino, the above
considerations are reasonable. The analysis may be difficult when
squarks become much heavier than gluinos.

So far we have demonstrated that there is finally no degeneracy
for the cases we studied. However, for realistic models, both the
squarks ($q'$) and gluinos ($g'$) could decay through many
different channels, leading to longer decay chain and more
final-state particles. This could change the jet multiplicity and
jet $P_T$ significantly, and it is not clear whether there are
degeneracies or not in these situations. If there were, one would
expect including more observables would again eliminate the
degeneracy. On the other hand, when LHC data are available, one
may be able to learn something about the decay topology or decay
branching ratios. Given that information, the potential model
dependence can be reduced. Finally, if there is no degeneracy,
then one can figure out the spin of gluino or squark by fitting to
the data.

\begin{table}[!t]
\begin{center}
  \begin{tabular}{|c||c|c|c|c|}
    \hline
    $m_{A}$ (GeV) & \multicolumn{4}{c|}{Same Spin Models ($m_{g'}$, $m_{q' }$) (GeV)}\\
    \cline{2-5}
    & (1000,1020) & (950,~1120) & (900,~1240) & (850,1400) \\ \hline\hline
     100  &  (1530,~9.6)  & (1657,~5.6)  & (1713,~7.2) & (1692,~5.8)    \\ \hline
     200  &  (1478,~8.9)  & (1585,~5.5)  & (1643,~7.0) & (1639,~6.1)   \\ \hline
     400  &  (1258,~10.3) & (1358,~4.3)  & (1377,~4.6) & (1381,~2.6)   \\ \hline
     600  &  (914,~13.6)  & (984,~0.8)  & (1015,~2.8) & (1037,~5.6)   \\ \hline
     800  &  (474,~20.4)  & (560,~15.3)  & (701,~19.8) & (961,~22.5)   \\ \hline
  \end{tabular}
  \end{center}
  \caption{This table shows 
  ($\overline H_T$, $\Delta S$) for each same spin model parameterized by three mass parameters
  $m_{g'}$, $m_{q'}$ and $m_A$ for Case II. Here $\Delta S$ is defined in Eq.(\ref{Eq:chisq}).
  In each column, a set of ($m_{g'}$, $m_{q'}$) is chosen to match the total cross section with that of the SUSY
  model, which is defined by $m_{\tilde g}=640$ GeV, $m_{\tilde q}=800$ GeV and $m_{\rm LSP}=100$ GeV.
  In each row, a mass $m_A$ is specified, which can be seen from the first column. For the SUSY model in the comparison,
  $\overline H_T=1161$ GeV, and the jet counts are given by \{7, 80, 387, 689, 834\} for jet number
  $n_j=1,2,3,4,5+$. The data indicate that $H_T$ and jet multiplicity of the same spin model cannot
  be matched to those in the SUSY model simultaneously}\label{Table:chisq2}
\end{table}

\section{Conclusions}\label{Conclusions}

We present a new method of determining the spins of TeV scale new
physics particles by utilizing production rate information.
Basically, if the mass and the production cross section are
measured, the spin is then determined. For many forms of new
physics that might be discovered, the method will work well.
Degeneracies may occur, in which case further analysis is needed,
which is provided. After clarifying the degeneracy in spin
measurement by using rate information in a single production
channel, we propose to combine information, such as jet counts and
$H_T$ distributions, from several channels. This method will be
particularly useful for an early determination of spin of the
colored new physics particles since we expect at least several of
them will be produced copiously. Such a method does not require
precise mass measurement. In particular, we consider two cases
where the SUSY models are characteristically different: $m_{\tilde
g}>m_{\tilde q}$ and $m_{\tilde g}<m_{\tilde q}$. For the first
case with a specific set of SUSY mass parameters ($m_{\tilde g}$,
$m_{\tilde q}$ and $m_{\rm LSP}$), we find that the jet
multiplicity cannot be well matched by the same spin model with
similar mass parameters ($m_{g'}$, $m_{q'}$ and $m_{A}$). In
addition, the $H_T$ distribution shows extra difference from that
of the SUSY model. For the second case, we find that the jet
multiplicity can be well matched by the same spin model. However,
the difference in the $H_T$ distribution can be used to
distinguish these two scenarios. The reason why this method works
relies on the fact that the production cross sections encode the
spin information of the SM partners in a nontrivial way, which
lead to signals which can be used to distinguish scenarios with
different spin assignments. In some cases, the pure rate signal
may be hard to see due to special mass splittings. However, the
kinematical differences are generically present and allow us to
distinguish these scenarios eventually.

Detailed study of spin determination using our method is certainly
needed. First of all, although we have imposed certain selection
cuts which mimic those would be used in Standard Model background
suppression, a precise evaluation of the effectiveness of our
method can only be obtained after carefully including  full
background and more realistic detector effects.

We have demonstrated our method using simple decay chains, which
could be the dominant ones in realistic models. At the same time,
more complicated decay patterns could well occur. On the one hand,
such decay channels, especially those with leptons, will greatly
enhance the discovery potential and signal to background ratio,
which could improve the sensitivity of of method. On the other
hand, they could introduce more model dependence. A detailed study
of disentangling those effect is certainly important and useful,
which will be discussed in follow-up papers.

\section{Acknowledgement}

We would like to thank Kyoungchul Kong, Bob McElrath, Aaron Pierce
and particularly Akin Wingerter for helpful comments and discussions. The
research of G.K. and J.S. is supported in part by the US
Department of Energy. A.A.P. was supported in part by th U.S.
National Science Foundation under CAREER Award PHY-0547794, and by
the U.S. Department of Energy under Contract DE-FG02-96ER41005.
The work of L.W. is supported by the National Science Foundation
under Grant No. 0243680 and the Department of Energy under grant
\# DE-FG02-90ER40542. Any opinions, findings, and conclusions or
recommendations expressed in this material are those of the
author(s) and do not necessarily reflect the views of the National
Science Foundation. J.S. and G.L.K. would like to thank the
Institute for Advanced Study - Princeton for hospitality, and
G.L.K. is grateful for funding in part by a grant from the Ambrose
Morrell Foundation. A.A.P. would like to thank Kavli Institute for
Theoretical Physics at UCSB for hospitality and acknowledge
support for his stay at KITP from the National Science Foundation
under Grant PHY05-51164.




\begin{thebibliography}{9}

\bibitem{Barr:2004ze}
  A.~J.~Barr,
  Phys.\ Lett.\  B {\bf 596}, 205 (2004)
  [arXiv:hep-ph/0405052].

\bibitem{Smillie:2005ar}
  J.~M.~Smillie and B.~R.~Webber,
  JHEP {\bf 0510}, 069 (2005)
  [arXiv:hep-ph/0507170].

\bibitem{Wang:2006hk}
  L.~T.~Wang and I.~Yavin,
  JHEP {\bf 0704}, 032 (2007)
  [arXiv:hep-ph/0605296].

\bibitem{Smillie:2006cd}
  J.~M.~Smillie,
  Eur.\ Phys.\ J.\  C {\bf 51}, 933 (2007)
  [arXiv:hep-ph/0609296].

\bibitem{Csaki:2007xm}
  C.~Csaki, J.~Heinonen and M.~Perelstein,
  JHEP {\bf 0710}, 107 (2007)
  [arXiv:0707.0014 [hep-ph]].

\bibitem{Meade:2006dw}
  P.~Meade and M.~Reece,
  Phys.\ Rev.\  D {\bf 74}, 015010 (2006)
  [arXiv:hep-ph/0601124].

\bibitem{Alves:2006df}
  A.~Alves, O.~Eboli and T.~Plehn,
  Phys.\ Rev.\  D {\bf 74}, 095010 (2006)
  [arXiv:hep-ph/0605067].


\bibitem{Bachacou:1999zb}
  H.~Bachacou, I.~Hinchliffe and F.~E.~Paige,
  Phys.\ Rev.\  D {\bf 62}, 015009 (2000)
  [arXiv:hep-ph/9907518].
\bibitem{Allanach:2000kt}
  B.~C.~Allanach, C.~G.~Lester, M.~A.~Parker and B.~R.~Webber,
  JHEP {\bf 0009}, 004 (2000)
  [arXiv:hep-ph/0007009].
\bibitem{Gjelsten:2004ki}
  B.~K.~Gjelsten, D.~J.~Miller and P.~Osland,
  JHEP {\bf 0412}, 003 (2004)
  [arXiv:hep-ph/0410303].

\bibitem{Gjelsten:2005aw}
  B.~K.~Gjelsten, D.~J.~Miller and P.~Osland,
  JHEP {\bf 0506}, 015 (2005)
  [arXiv:hep-ph/0501033].

\bibitem{Lester:2006yw}
  C.~G.~Lester,
  Phys.\ Lett.\  B {\bf 655}, 39 (2007)
  [arXiv:hep-ph/0603171].

\bibitem{Lester:2005je}
  C.~G.~Lester, M.~A.~Parker and M.~J.~White,
  JHEP {\bf 0601}, 080 (2006)
  [arXiv:hep-ph/0508143].

\bibitem{Ross:2007rm}
  G.~G.~Ross and M.~Serna,
  arXiv:0712.0943 [hep-ph].

\bibitem{Barr:2007hy}
  A.~J.~Barr, B.~Gripaios and C.~G.~Lester,
  arXiv:0711.4008 [hep-ph].

\bibitem{Cho:2007dh}
  W.~S.~Cho, K.~Choi, Y.~G.~Kim and C.~B.~Park,
  arXiv:0711.4526 [hep-ph].

\bibitem{Lester:2007fq}
  C.~Lester and A.~Barr,
  arXiv:0708.1028 [hep-ph].

\bibitem{Cheng:2007xv}
  H.~C.~Cheng, J.~F.~Gunion, Z.~Han, G.~Marandella and B.~McElrath,
  arXiv:0707.0030 [hep-ph].

\bibitem{Cheng:2008mg}
  H.~C.~Cheng, D.~Engelhardt, J.~F.~Gunion, Z.~Han and B.~McElrath,
  arXiv:0802.4290 [hep-ph].



\bibitem{Tovey:2000wk}
  D.~R.~Tovey,
  Phys.\ Lett.\  B {\bf 498}, 1 (2001)
  [arXiv:hep-ph/0006276].

\bibitem{Hinchliffe:1996iu}
  I.~Hinchliffe, F.~E.~Paige, M.~D.~Shapiro, J.~Soderqvist and W.~Yao,
  Phys.\ Rev.\  D {\bf 55}, 5520 (1997)
  [arXiv:hep-ph/9610544].

\bibitem{Tevatron-top-rate-cdf}
The CDF Collaboration. Conf. Note 8148.\\
http://www-cdf.fnal.gov/physics/new/top/public$\_$xsection.html

\bibitem{Tevatron-top-rate-d0}
The D0 Collaboration. D0 Note 5591-CONF.\\
http://www-d0.fnal.gov/Run2Physics/WWW/results/prelim/TOP/T64/

\bibitem{Tevatron-top-mass-cdf}
The CDF Collaboration. Conf. Note 9214. \\
http://www-cdf.fnal.gov/physics/new/top/public$\_$mass.html


\bibitem{Tevatron-top-mass-d0}
The D0 Collaboration. D0 Note 5498-CONF. \\
http://www-d0.fnal.gov/Run2Physics/WWW/results/prelim/TOP/T63/

\bibitem{Datta:2005vx}
  A.~Datta, G.~L.~Kane and M.~Toharia,
  arXiv:hep-ph/0510204.


\bibitem{Appelquist:2000nn}
  T.~Appelquist, H.~C.~Cheng and B.~A.~Dobrescu,
  Phys.\ Rev.\  D {\bf 64}, 035002 (2001)
  [arXiv:hep-ph/0012100].

\bibitem{Cheng:2003ju}
  H.~C.~Cheng and I.~Low,
  JHEP {\bf 0309}, 051 (2003)
  [arXiv:hep-ph/0308199].

\bibitem{Cheng:2004yc}
  H.~C.~Cheng and I.~Low,
  JHEP {\bf 0408}, 061 (2004)
  [arXiv:hep-ph/0405243].

\bibitem{nlo_gluino_squark}
  W.~Beenakker, R.~Hopker, M.~Spira and P.~M.~Zerwas,
  Nucl.\ Phys.\  B {\bf 492}, 51 (1997)
  [arXiv:hep-ph/9610490].
\bibitem{Plehn:2005cq}
  T.~Plehn, D.~Rainwater and P.~Skands,
  Phys.\ Lett.\  B {\bf 645}, 217 (2007)
  [arXiv:hep-ph/0510144].

\bibitem{nlo_stop}
  W.~Beenakker, M.~Kramer, T.~Plehn, M.~Spira and P.~M.~Zerwas,
  Nucl.\ Phys.\  B {\bf 515}, 3 (1998)
  [arXiv:hep-ph/9710451].

\bibitem{OctetScalars}  S.~I.~Bityukov and N.~V.~Krasnikov,
Mod.\ Phys.\ Lett.\ A \textbf{12}, 2011 (1997)  [arXiv:hep-ph/9705338];
A.~V.~Manohar and M.~B.~Wise,
Phys.\ Rev.\ D \textbf{74}, 035009 (2006)  [arXiv:hep-ph/0606172].
M.~I.~Gresham and M.~B.~Wise,
Phys.\ Rev.\  D {\bf 76}, 075003 (2007)
[arXiv:0706.0909[hep-ph]].

\bibitem{OctetVectors}
C.~Macesanu, C.~D.~McMullen and S.~Nandi,
Phys.\ Rev.\  D {\bf 66}, 015009 (2002) [arXiv:hep-ph/0201300].


\bibitem{Dawson:1983fw}  S.~Dawson, E.~Eichten and C.~Quigg,
Phys.\ Rev.\ D \textbf{31}, 1581 (1985).  


\bibitem{Beenakker:1996ch}  W.~Beenakker, R.~Hopker, M.~Spira and
P.~M.~Zerwas,  
Nucl.\ Phys.\ B \textbf{492}, 51 (1997)  [arXiv:hep-ph/9610490].

\bibitem{Yao:2006px}
  W.~M.~Yao {\it et al.}  [Particle Data Group],
  J.\ Phys.\ G {\bf 33} (2006) 1.

\bibitem{Cheng:2005as}
  H.~C.~Cheng, I.~Low and L.~T.~Wang,
  Phys.\ Rev.\  D {\bf 74}, 055001 (2006)
  [arXiv:hep-ph/0510225].

\bibitem{ArkaniHamed:2005px}
  N.~Arkani-Hamed, G.~L.~Kane, J.~Thaler and L.~T.~Wang,
  JHEP {\bf 0608}, 070 (2006)
  [arXiv:hep-ph/0512190].

\bibitem{madgraph}
  F.~Maltoni and T.~Stelzer,
  JHEP {\bf 0302}, 027 (2003)
  [arXiv:hep-ph/0208156].
  J.~Alwall {\it et al.},
  JHEP {\bf 0709}, 028 (2007)
  [arXiv:0706.2334 [hep-ph]].
\bibitem{cteq6}
  J.~Pumplin, D.~R.~Stump, J.~Huston, H.~L.~Lai, P.~Nadolsky and W.~K.~Tung,
  JHEP {\bf 0207}, 012 (2002)
  [arXiv:hep-ph/0201195].
\bibitem{pythia6}
  T.~Sjostrand, S.~Mrenna and P.~Skands,
  JHEP {\bf 0605}, 026 (2006)
  [arXiv:hep-ph/0603175].
\bibitem{pgs4}
http://www.physics.ucdavis.edu/~conway/research/software/pgs/pgs4-general.htm

\bibitem{Meade:2007js}
  P.~Meade and M.~Reece,
  arXiv:hep-ph/0703031.

\end{thebibliography}
\end{document}